\documentclass[fleqn,12pt,twoside]{article}
\usepackage{espcrc1}

\usepackage{graphicx}
\def\s2w{\sin^2\theta_W}
       
\begin{document}

% declarations for front matter
\title{Parity Violation with Electrons and Hadrons}
\author{E.J.~Beise\address{Dept.~of Physics, University of Maryland, 
College Park, MD, USA}}

% typeset front matter
\maketitle

\begin{abstract}

A key question in  understanding the structure of
nucleons involves the role of sea quarks in their ground state electromagnetic
properties such as charge and magnetism. Parity-violating
electron scattering, when combined with determination of nucleon electromagnetic
form factors from parity-conserving $e$-$N$ scattering, provides another
degree of freedom to separately determine the up, down and strange quark
contributions to nucleon electromagnetic structure. Strange quarks are unique in
that they are exclusively in the nucleon's sea. A program of experiments
using parity violating electron scattering 
has been underway for approximately a decade, and results are beginning to 
emerge. This paper is a brief overview of the various experiments and their
results to date along with a short-term outlook of what can be anticipated
from experiments in the next few years.

\end{abstract}

\section{Introduction}

In 1988, Kaplan and Manohar~\cite{Kap88} showed that information about
the contribution of sea quarks to ground state nucleon
properties, such as spin, charge and magnetic moments, could be learned
using neutral weak probes of the nucleon such as neutrino-nucleon
scattering. Soon after, McKeown~\cite{BMcK89} and
Beck~\cite{Bec89} outlined a possible program of experiments in
parity-violating electron scattering that, when combined with existing
measurements of nucleon electromagnetic form factors, would
allow the identification of possible strange quark contributions
to the proton's charge and magnetism. Experiments were proposed
at the MIT-Bates Laboratory, Jefferson Lab, and the Mainz Microtron 
to accomplish this goal.  In the decade or so since, many theoretical 
models of the strange quark components of these neutral weak matrix elements,
have appeared in the literature, and the first quantitative information from
experiments is becoming available. Here I will provide a summary of 
the recent measurements, along with expected progress in the short term.

While the expectation is that strange quark contributions 
should be small, they occupy a special place in nucleon structure because
their presence would be exclusively in the nucleon's $\overline{q}q$ sea.
Evidence to date suggests that they have a sizeable contribution
to the nucleon's unpolarized quark momentum distribution
in the nucleon~\cite{NuT00}, as well as to the nucleon's 
mass~\cite{Ols00}, although the latter has some degree 
of uncertainty due to both experimental and theoretical extrapolations
required to obtain the result. A decade of precise spin-dependent deep-inelastic
scattering experiments has led to the conclusion that strange quarks
contribute significantly to the (small) fraction of the proton's spin 
carried by quark spins~\cite{Lea02}, although again assumptions about
SU(3) symmetry are required in order to extract a result.

Parity-violating electron scattering is primarily 
sensitive to the matrix element $\overline{s}\gamma_{\mu} s$, which
provides information about the $\overline{s}s$ contributions to the
nucleon's charge and magnetization distributions. The neutral weak
nucleon current as probed through PV $e$-$N$ scattering is
\begin{equation}
J_{\mu}^{NC} \equiv
\langle N\vert\hat{J}_{\mu}^{NC} \vert N\rangle
= {\overline U}\left[ \gamma_{\mu} F_1^{Z}(Q^2)
+ i\sigma_{\mu\nu}q^{\nu}\frac{F_2^{Z}(Q^2)}{2M} +
\gamma_{\mu}\gamma_5 G_A^{Z}(Q^2) \right] U \, ,
\end{equation}
where $F_{1,2}^{Z}$ are the neutral weak equivalents of the nucleon's
Dirac and Pauli form factors $F_{1,2}$.  At low momentum transfer, 
$F_1$ and $F_2$ are more often expressed as the Sachs form factors $G_E = 
F_1 + \frac{Q^2}{4M^2}F_2$ and $G_M=F_1+F_2$, which can be 
directly related to the nucleon's charge and magnetization
distributions, respectively, and which have the normalizations 
$G_E^p(0)=1$, $G_E^n(0)=0$, $G_M^p(0)=\mu_p$, $G_M^n(0)=\mu_n$.
Corresponding definitions can be made for the vector weak 
form factors $F_{1,2}^{Z}$. Because the NW form factors are
derived from the same matrix element $\overline{q}\gamma_\mu q$
as their EM counterparts, they can be re-expressed in terms of
the measured EM form factors, with an explicit remainder coming
from strange quarks, assuming only that neutrons and protons differ
by an interchange of $u$ and $d$ quarks. 
\begin{equation}
\label{eq:gemz}
G_{E,M}^Z = \left(1-4\s2w\right)\left(1+R_V^p\right)G_{E,M}^{\gamma ,p} 
-\left(1+R_V^n\right) G_{E,M}^{\gamma ,n} - G_{E,M}^s  \, .
\end{equation}
The $G_{E,M}^Z$ thus provide a third degree of freedom to 
disentangle the flavor structure of the proton's charge and magnetism.
The radiative corrections $R_V^{p,n}$ represent contributions from 
higher order processes and have been computed~\cite{Mus94} 
to be $R_V^p=-0.053\pm0.033$ and $R_V^n=-0.0143\pm0.0004$. 

The axial form factor $G_A^{Z}$ is related to the same matrix 
element as that defining the nucleon's spin, $\overline{s}\gamma_\mu\gamma_5 s$,
and its isoscalar component
explicitly contains the $s$-quark contribution, $\Delta s$. Its isovector 
 $(T=1)$ component, to which PV $e$-$p$ scattering is primarily sensitive,
can be expressed in terms of the neutron $\beta$-decay constant
$(g_A/g_V)=-1.2670\pm0.0035$, but also contains higher order corrections that
can come from, for example, an electromagnetic $e$-$p$ interaction 
coupled with a weak exchange between quarks, or from $\gamma$-$Z$ 
box diagrams. The effective axial form factor is 
\begin{equation}
G_A^e = -\tau_3 G_A(Q^2) + \Delta s + \eta F_A + R_e\, ,
\end{equation}
where $\tau_3$=+($-$)1 for protons(neutrons), $G_A(Q^2)=(g_A/g_V)/(1+Q^2/M_A^2)^2$,
and $(\eta F_A + R_e)$ are due to the higher order terms. 

Experimentally, parity-violating $e$-$p$ scattering results in an 
asymmetry in the detected yield for a longitudinally polarized 
beam on an unpolarized target:
\begin{equation}
\label{eq:pvee}
A_{PV} = \frac{d\sigma_R - d\sigma_L}{d\sigma_R + d\sigma_L}
 = -\frac{G_FQ^2}{4\pi\alpha\sqrt{2}}
\frac{A_E + A_M + A_A}{\left[\varepsilon \left(G_E^\gamma\right)^2 +
\tau \left(G_M^\gamma\right)^2\right]}
\end{equation}
where
\begin{eqnarray}
A_E &=& \varepsilon G_E^Z G_E^\gamma\, ,\;\;\; A_M = \tau G_M^Z G_M^\gamma \, 
,\nonumber \\
A_A &=& -\left(1-4\s2w\right)
\sqrt{\tau\left(1+\tau\right)\left(1-\varepsilon^2\right)} G_A^e G_M^\gamma\, ,
\end{eqnarray}
$\tau$ and $\varepsilon$ are kinematic factors and the $\gamma$ index 
refers to nucleon EM form factors. A single measurement typically
involves at least two of the above terms, so a complete separation of
$G_E^Z$, $G_M^Z$ and $G_A^e$ involves at least three experiments.  
Quasielastic scattering from deuterium can be used to constrain $G_A^e(T=1)$ 
since it carries similar sensitivity to the axial term but is relatively
insensitive to the strange vector form factors. A program
of experiments has been carried out at MIT-Bates and Jefferson Laboratories
and new experiments are underway at JLab and also at the Mainz Microtron.

\begin{figure}
\begin{center}
\includegraphics[width=3.0in,angle=90]{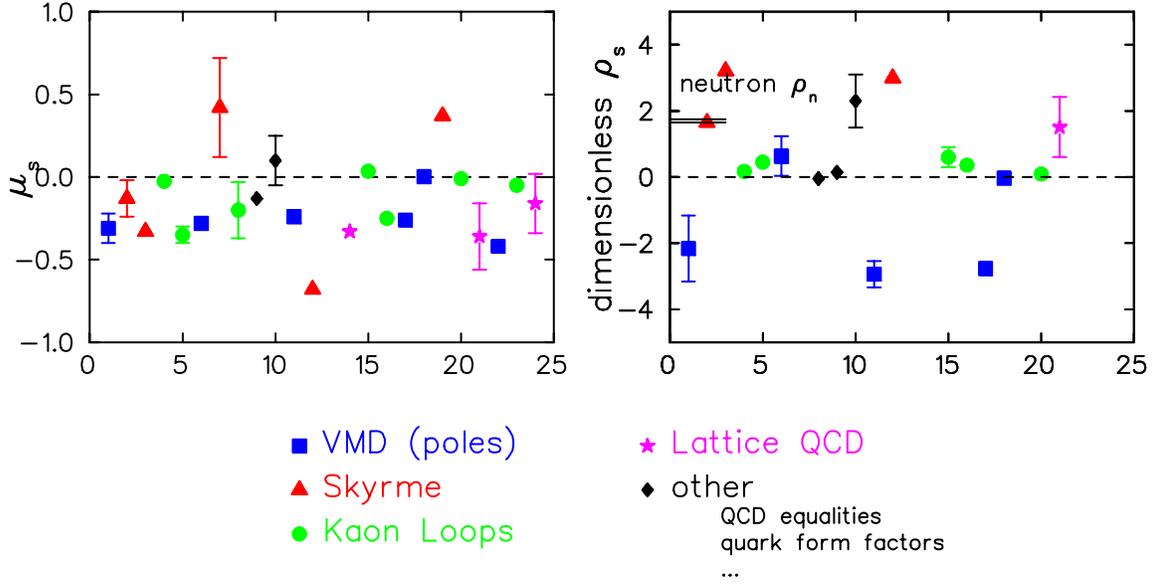}
\caption{Graphical summary of the various types model predictions for $\mu_s$ and
the electric strangeness radius parameter $\rho_s$. The horizontal
axis refers to a particular calculation: see~\protect{\cite{BeH01}} for 
a recent summary. For scale, the equivalent
radius parameter for the neutron's charge form factor is also shown.}
\end{center}
\label{fig:models}
\end{figure}

On the theoretical front, a wide variety of models have been used to
estimate the magnitude of $G_E^s$ and $G_M^s$, including a few predictions
for their behavior with $Q^2$. At $Q^2$=0, $G_E^s$ is constrained to 
be 0 since the proton has no net strangeness, but $G_M^s$ is not so
constrained and is defined through the expression
\begin{equation}
\mu_p = \frac{2}{3}\mu_u - \frac{1}{3}\mu_d - \frac{1}{3}\mu_s\, .
\end{equation}
The low energy behavior of each is characterized by a radius parameter
that can be written in a dimensionless form as 
$\rho =4M_N^2 \left(dG/dQ^2\right)\vert_{Q^2=0}$. The wide variety of
predictions precludes extensive discussion here, but reviews
can be found in the literature~\cite{BeH01}. Figure~\ref{fig:models}
shows a sampling of many of the models: a notable feature
is that while many predict $\mu_s$ to be $\sim -0.3$, predictions for the
electric strangeness $\rho_s$ vary widely and do not even agree on the sign.

It is of interest to note one recent calculation~\cite{Hem98}, where $G_M^s$ and
$G_E^2$ were analyzed within the framework of chiral perturbation theory.
It was first thought that the slope of $G_M^s$ could be determined analytically,
and with constraints coming from the first results from SAMPLE and HAPPEX,
limits on the $Q^2$ behavior could be predicted~\cite{Hem99}, which
resulted in opposite signs for the two form factors. 
It was, however, recently shown in~\cite{Ham03} that the slope 
of $G_M^s$ is sensitive to an unknown low energy constant that enters at 
$O(p^4)$ due cancellations at lower order, so both the magnitude and 
sign of $G_M^s$ at low $Q^2$ are still unconstrained by theory.

\section{SAMPLE at MIT-BATES}

In the SAMPLE experiment~\cite{Spa00,Has00}, PV electron scattering 
was measured in the backward direction, from both hydrogen and deuterium targets, 
in order to determine $G_M^s$ and $G_A^e(T=1)$ at
$Q^2=0.1$~(GeV/c)$^2$. A 200~MeV, 40~$\mu$A polarized beam was scattered from
a 40~cm liquid hydrogen target. Cerenkov light from scattered electrons was
detected in an array of ten mirror-phototube pairs arranged symmetrically
about the beam axis covering angles between 130$^\circ$ and 170$^\circ$. 
The scattered electron rate was integrated over the
25~$\mu$sec beam pulse and sorted by beam helicity state, which was flipped
pseudo-randomly at 600~Hz. Results from the 200~MeV running were published
in~\cite{Spa00} and~\cite{Has00}, where the latter included data from
quasielastic scattering from deuterium. It was found that
while $G_M^s$ is likely small, there was an approximately 1.5$\sigma$ 
discrepancy between the extracted value of $G_A^e(T=1)$ and that expected
assuming a value for the weak radiative corrections as computed by 
Zhu~{\it et al.}~\cite{Zhu00}. This led to theoretical investigations of
the nuclear contributions in the deuterium data as well as a more detailed
look at the extraction of $G_A^e(T=1)$ from the data.

In a simple static approximation, the PV asymmetry in deuterium is an 
incoherent sum of that of the neutron 
and proton, and the contributions from $G_M^s$ largely cancel, and the 
sensitivity to $G_A^e(T=1)$ is approximately the same as for a proton target.
Nuclear corrections can potentially modify the asymmetry, both through
parity-conserving~\cite{Dia01}, and parity-violating~\cite{Liu03,Schi03} 
two-body effects. At the SAMPLE kinematics the PC terms modify the 
asymmetry by 1-3\%, and the PV terms are negligible. The SAMPLE apparatus
also has contributions from threshold breakup and elastic $e$-$d$ scattering 
which modify the asymmetry by a few percent.

\begin{figure}
\begin{center}
\includegraphics[height=3.0in]{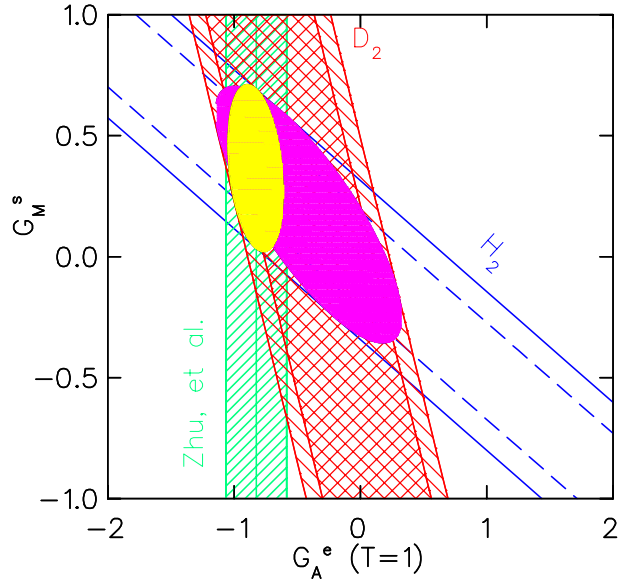}
\caption{Updated results from the 200~MeV SAMPLE data, which resulted
in better agreement with the theoretical expectation~\protect{\cite{Zhu00}} 
for the axial form factor.}
\end{center}
\label{fig:sample}
\end{figure}

An improved analysis of the SAMPLE deuterium data now brings the 
extracted value of $G_A^e(T=1)$ into reasonable agreement with~\cite{Zhu00},
as shown in Figure~\ref{fig:sample}, but has a relatively small
impact on $G_M^s$. The new results include a complete 
GEANT model of the detector, a revision to the electromagnetic 
radiative corrections and a dilution correction for coherent $\pi^0$ production 
in the experimental yield which was previously neglected. Furthermore, 
the calculation of~\cite{Schi03} was used to model the physics 
asymmetry. These combined theoretical and experimental efforts lead to
better confidence that the higher order contributions to 
$G_A^e$ are now under control.  Results from the third SAMPLE measurement, 
at lower momentum transfer, also agree with expectations from theory~\cite{Ito03}.
The hydrogen results were also revised~\cite{Spa01}, resulting an experimental
asymmetry of, after all dilution corrections, 
$A_{exp} = -5.61 \pm 0.68 \pm 0.88$~ppm. Combining this result with 
the theoretical value of $G_A^e$ results~\cite{Bei03} results in the 
more upright ellipse in Figure~\ref{fig:sample}. While this result is 
consistent with little or no strange quark effects, it suggests a
positive value for $G_M^s$ whereas most calculations predict 
a $Q^2=0$ value near $\mu_s\sim-0.3$.

\section{HAPPEX at JLab}

The first measurements of parity-violating electron scattering at Jefferson
Laboratory~\cite{Ani00} 
were carried out by the HAPPEX collaboration, who used a 3.3~GeV
polarized beam on a 15~cm hydrogen target and detected the scattered electrons
using the pair of high resolution spectrometers (HRS) in Hall A at 12.5$^\circ$.
The measured asymmetry, at $Q^2$=0.48~(GeV/c)$^2$, is sensitive to the combination
$G_E^s + 0.39G_M^s$. The counting rate of approximately 1~MHz per spectrometer
required the use of integrating techniques, and a set of Pb-scintillator
total absorption counters was used instead of the standard tracking 
detector package. The HAPPEX experiment 
was the first to use a strained GaAs crystal in the polarized electron source,
which produces beam polarization in excess of 70\%. Such sources can have
an analyzing power for linear polarization in the incident laser light that 
could potentially result in significant helicity correlated position differences
of the beam on the experimental target. Such effects were, however, kept to
a negligible level by insertion of a rotatable half-wave plate in the 
laser beam and with a feedback system nulling any helicity-correlated intensity
asymmetry.

The experimentally determined asymmetry from the HAPPEX experiment is 
$A_{exp}=-15.05\pm 0.98\pm 0.56$~ppm, corresponding to $G_E^s+0.39G_M^s =
0.025 \pm 0.020 \pm 0.014$ where the latter uncertainty is due to knowledge
of the nucleon EM form factors. HAPPEX thus for the most part precludes
the parameter space in which $G_E^s$ and $G_M^s$ have the same sign.

The future program for HAPPEX includes a forward angle measurement at
$Q^2$=0.1 (GeV/c)$^2$ on hydrogen~\cite{Kum99}, as well as the first measurement of
the PV asymmetry in elastic electron scattering from helium~\cite{Arm00}. 
Due to the fact that $^4$He is a spin-0, isospin-0, target, only a single weak
form factor exists and it can be directly related to $G_E^s$ with a good
model of the $^4$He nucleus. Theoretical expectations are that contributions
to the asymmetry from many-body effects in the helium are 
negligible~\cite{MusD93} at low momentum transfer. The combined measurements,
or the new hydrogen measurement combined with the SAMPLE result, will
result in a determination of $G_E^s$.

\section{PVA4 at Mainz}

The PVA4 collaboration at Mainz has taken a different experimental approach,
with a detector with sufficient segmentation and specialized electronics 
that counting the scattered particles is feasible despite high rates. A 20~$\mu$A beam
of polarized electrons was incident on a 10~cm target. Scattered electrons
are detected at 35$^\circ$ with a PbF$_2$ {\v C}erenkov shower calorimeter.
The detector design includes 1022 PbF$_2$ crystals arranged in 7 rings, and
processed in 3x3 modules with self-triggering and histogramming electronics:
for the first two measurements approximately half of the detector was instrumented.
The energy resolution of the detectors must be sufficient to separate the 
10~MHz of elastically scattered electrons from the 90~MHz of inelastic electrons 
coming from threshold pion and resonance production. The achieved 
energy resolution was 4\%/$\sqrt{E}$~\cite{Wie03}.

The first PVA4 measurement was at a beam energy of 855~MeV, corresponding to
$Q^2$=0.23~(GeV/c)$^2$ and a sensitivity to the combination $G_E^s+0.22 G_M^s$.
The experimental asymmetry after all dilution corrections is 
$A_{exp} = -5.6 \pm 0.6 \pm 0.2$~ppm, corresponding to an approximately 1$\sigma$
deviation of the asymmetry from that expected with no strange quarks, but 
again hinting that $G_E^s$ and $G_M^s$ are either both small or have 
opposite sign. Additional data have already been taken, the run concluding in June 2003, 
at 570~MeV beam energy, corresponding to $Q^2$=0.1~(GeV/c)$^2$. Again by
combining these data with the results from SAMPLE will allow the first
experimental limits on $G_E^s$. Future plans involve reversing the detector
for backward angle measurements at $Q^2$=0.23 and 0.48~(GeV/c)$^2$ to combine
with the existing HAPPEX and PVA4 data. 

\begin{figure}
\begin{center}
\includegraphics[height=3.0in]{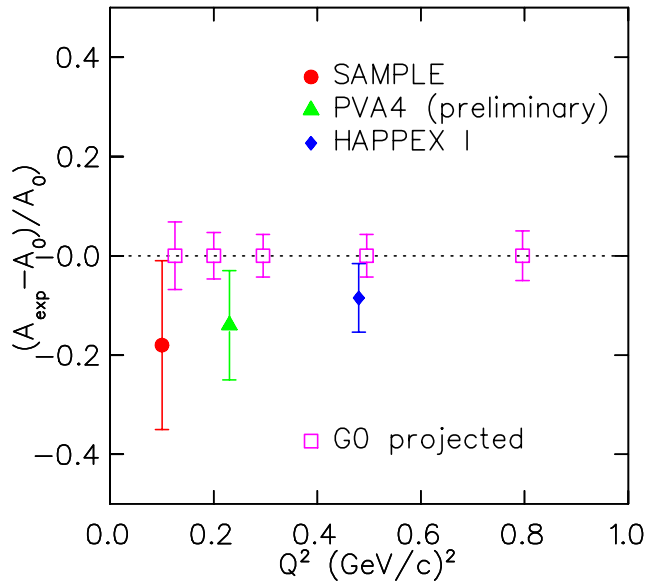}
\caption{Summary of existing measurements of the PV asymmetry in elastic
$e$-$p$ scattering, shown as the fractional deviation of the measurement from
the asymmetry expected if $G_M^s$=$G_E^s$=0 at all momentum transfers.
Overlaid are the expected uncertainties coming from the first phase of
the G0 experiment, scheduled to run in early 2004.}
\end{center}
\label{fig:asym-dev}
\end{figure}

Although the kinematic sensitivities of each of the experiments is somewhat different, 
the results from each of the three above $e$-$p$ measurements can be shown, as in 
Figure~\ref{fig:asym-dev}, as a deviation from the asymmetry expected with no 
strange quarks. While any single measurement is consistent with little or
no strange quark contribution, the trend in all three experiments suggests
an $s$-quark contribution that is slightly positive.

It should be noted that, in addition to the parity-violation results, both the 
PVA4 and SAMPLE experiments have measured a ``beam spin asymmetry'' resulting
from scattering from a purely transversely polarized beam~\cite{Maa03,Wel01}. 
To lowest order, the asymmetry, which results in a variation of the cross section 
in azimuthal
angle with respect to the beam axis, is to lowest order the result of two-photon
processes. Such processes have recently become of interest because they
may help explain the discrepancy in the determination of the proton charge
form factor at high momentum transfer from polarization and cross section
data~\cite{Blu03}, and they are related to the Virtual Compton Scattering process which
provide information about nucleon polarizabilities~\cite{Van03}.

\section{G0 at JLab}

The G0 experiment at JLab is a dedicated to determining $G_E^s$, $G_M^s$ 
and $G_A^e$ from a single experimental apparatus over a broad $Q^2$ range. 
The detector consists of a superconducting toroidal spectrometer
with an array of scintillators along the focal plane to determine the 
PV asymmetry at both forward and backward scattered electron angles.
Polarized electrons are scattered from a 20~cm liquid hydrogen target.
In the forward configuration, the recoil protons are detected and sorted
by $Q^2$ covering the range $0.1 < Q^2 < 1.0$~(GeV/c)$^2$. In the
backward configuration the apparatus is reversed and electrons scattered
at 108$^\circ$ will be detected with three dedicated magnet settings 
corresponding to $Q^2$ = 0.25, 0.5 and 0.8~(GeV/c)$^2$. In the forward
mode, the protons are identified and separated from pions via time-of-flight 
from the target to detector, which is about 20~ns. In the backward mode 
the detector package requires an augmentation with another array of
scintillators (Cryostat Exit Detectors) and aerogel {\v C}erenkov counters to
eliminate negative pions from the trigger. Data will also be acquired with 
a deuterium target to experimentally determine the $Q^2$-dependence of $G_A^e$.

The first engineering run of the experiment, in its forward mode, 
was carried out in late 2002 during which a few days of asymmetry data were 
collected. While the statistical uncertainties are too large to draw 
physics conclusions from these data, the measured asymmetry of approximately 
5~ppm is consistent with expectation, has a reasonable
$Q^2$ behavior, and reverses appropriately with manual reversal of the 
sign of the beam polarization.
After a second engineering run, the G0 collaboration will carry out its 
forward mode physics measurement. While a definitive determination of the relative
signs and/or magnitude of $G_M^s$ and $G_E^s$ will require the backward angle
measurements as well, this first set of data will both extend the kinematic 
reach and improve the precision of the data shown in Figure~\ref{fig:asym-dev}.  
Additional details of the G0 experiment can be found in~\cite{Pit03} 
and~\cite{Bat03}.

%\begin{figure}
%\begin{center}
%\includegraphics[width=3in,angle=-90]{combined_na_fr_per_hwp.ps}
%\caption{Measured asymmetries as a function of detector number,
%from the first engineering run of the G0
%experiment, coming from a few percent of the amount of data
%expected in the first physics run planned for early 2004. The two sets
%of data correspond to manual insertion of a half-wave plate in the
%injector laser which reverses the sign of the beam polarization. The data
%are uncorrected for beam polarization, background contributions, or helicity
%correlations in the beam.}
%\end{center}
%\label{fig:g0results}
%\end{figure}

Improvements in polarized beam technology in the last decade
have made precise measurements of parity-violating electron 
scattering possible, and the next generation of
experiments, which move beyond studies of hadron structure to other physics
are now being considered. One direction is to use parity-violating electron
scattering as a precise probe of neutron distributions in heavy nuclei
~\cite{Sou03}, which may have relevance in understanding the structure of
neutron stars. Another direction is to carry out precision
tests of the standard model at relatively low momentum transfer where sensitivity
to additional $Z$-bosons, for example, is greatly enhanced. The latter is
of particular interest in light of the recent results from the NuTeV collaboration
in which a 3$\sigma$ deviation from expectation in $\sin^2\theta_W$ was 
measured~\cite{Zel02}.
The latter direction is being pursued by the QWEAK collaboration at JLab, and
further details can be found in~\cite{Pit03} in these proceedings.

\section{Summary}

Since the earliest measurements of parity-violating electron scattering at 
SLAC~\cite{Pres78} in which the weak mixing angle was first measured, the 
basic techniques of parity-violating electron scattering have remained 
more or less the same. But the achievable precision has greatly improved
as a result of new high intensity electron beams, advances in polarized 
beam technology, and technical advances in the feedback and laser systems
that are needed for polarized beam delivery. New physics directions also 
emerged and the results are beginning to become available from the first
experiments to use PV electron scattering as a probe of hadron structure. 
We can look forward to more new results in the near future with the next
phases of HAPPEX and PVA4 and the first results from the G0 experiment, all
of which should further our understanding of the role of sea quarks in
the nucleon's ground state structure.

The author is supported by NSF contract~PHY-0140010. Much of
the work discussed in this paper has been supported both by 
the National Science Foundation and the U.S.~Dept.~of Energy.


\begin{thebibliography}{9}
\bibitem{Kap88} D.~Kaplan and A.V.~Manohar, Nucl.~Phys.~{\bf B310}, 527 (1988).
\bibitem{BMcK89} R.D.~McKeown, Phys.~Lett.~{\bf B219}, 140 (1989).
\bibitem{Bec89} D.H.~Beck, Phys.~Rev.~{\bf D39}, 3248 (1989).
\bibitem{NuT00}  T.~Adams {\it et al.}, NuTeV Collaboration, hep-ex/9906038;
V.~Barone, C.~Pascaud, and F.~Zomer, hep-ph/0004268. See also 
M.~Goncharov {\it et al.}, Phys.~Rev.~{\bf D 64}, 112006 (2001).
\bibitem{Ols00} M.~Olsson, Phys.~Lett.~{\bf B482}, 50 (2000).
\bibitem{Lea02} E.~Leader, A.V.~Siderov, and D.B.~Stamerov, 
  Eur.~Phys.~J.~{\bf C 23},479 (2002).
\bibitem{Mus94} M.~J.~Musolf, {\it et al.}, Phys.~Rep.~{\bf 239}, 1 (1994).
\bibitem{Spa00} D.T.~Spayde {\it et al.}, Phys.~Rev.~Lett.~{\bf 84}, 1106 (2000).
\bibitem{Has00} R.~Hasty {\it et al.}, Science {\bf 290}, 2117 (2000).
\bibitem{BeH01} See  D.H.~Beck and B.R.~Holstein, 
  Int.~J.~Mod.~Phys.~{\bf E10}, 1 (2001), for a recent review.
\bibitem{Hem98} T.~R.~Hemmert, U.-G.~Meissner, and S.~Steininger,
  Phys.~Lett.~{\bf B 437}, 184 (1998).
\bibitem{Hem99} T.~R.~Hemmert, B.~Kubis, and U.-G.~Meissner,
  Phys.~Rev.~{\bf C 60}, 045501 (1999).
\bibitem{Ham03} H.-W.~Hammer, S.J.~Puglia, M.J.~Ramsey-Musolf, and S.-L.~Zhu,
  Phys.~Lett.~{\bf B 562}, 208 (2003).
\bibitem{Zhu00} S.-L.~Zhu, S.J.~Puglia, B.R.~Holstein, and M.J.~Ramsey-Musolf,
  Phys.~Rev.~{\bf D 62}, 033008 (2000).
\bibitem{Dia01}L.~Diaconescu, R.~Schiavilla, and U.~van Kolck, 
  Phys.~Rev.~{\bf C 63}, 044007 (2001).
\bibitem{Liu03} C.-P.~Liu, G.~Pr\'{e}zeau, and M.J.~Ramsey-Musolf,
  Phys.~Rev.~{\bf C 67}, 035501 (2003).
\bibitem{Schi03}R.~Schiavilla, J.~Carlson and M.~Paris,
  Phys.~Rev.~{\bf C 67}, 032501(R) (2003).
\bibitem{Ito03}T.~Ito {\it et al.}, paper in progress.
\bibitem{Spa01} D.T.~Spayde, Univ.~Md.~Ph.D. thesis, May 2001, unpublished,
   but can be found at {\it http://www.physics.umd.edu/enp/theses/}.
\bibitem{Bei03}E.J.~Beise {\it et al.}, paper in progress.
\bibitem{Ani00}K.~Aniol, {\it et al.}, Phys.~Lett~{\bf B 509}, 211 (2001).
   See also K.~Aniol, {\it et al.}, Phys.~Rev.~Lett.~{\bf 82}, 1096 (1999.)
\bibitem{Kum99} JLAB experiment E99--115, K.~Kumar and D.~Lhuillier, contacts.
\bibitem{Arm00} JLAB experiment E00--114, D.~Armstrong and R.~Michaels,
   contacts.
\bibitem{MusD93} M.J.~Musolf and T.W.~Donnelly,
   Phys.~Lett.~{\bf B 318}, 263 (1993).
\bibitem{Wie03} J.~van de Wiele and M.~Morlet, for the A4 collaboration,
   Czech.~Jour.~Phys.~{\bf 53}, A1 (2003). An updated experimental
   result was provided for this conference by F.~Maas, private communication.
\bibitem{Maa03} F.~Maas, private communication, and talk given at the ECT, Trento,
   April 2003.
\bibitem{Wel01} S.P.~Wells {\it et al.},  Phys.~Rev.~{\bf C 63}, 064001 (2001).
\bibitem{Blu03} P.~Blunden, W.~Melnitchouk, and J.A.~Tjon, nucl-th/0306076 (2003).
\bibitem{Van03} M.~Vanderhaeghen and S.P.~Wells, private communication.
\bibitem{Pit03} M.~Pitt, these proceedings.
\bibitem{Bat03} G.~Batigne contribution to the ``Fourth International Conference 
on Perspectives in Hadronic Physics'', to appear in Euro.~Phys.~Jour.~{\bf A}, 2003.
\bibitem{Sou03} JLab experiment E99-012, R.~Michaels and P.~Souder, contacts.
\bibitem{Zel02} G.P.~Zeller {\it et al.},  Phys.~Rev.~Lett.~{\bf 88}, 091802 (2002).
   See also {\it ibid.}, Phys.~Rev.~{\bf D 65}, 111103 (2002). 
\bibitem{Pres78} C.Y.~Prescott {\it et al.}, Phys.~Lett.~{\bf B 77}, 347 (1978).

\end{thebibliography}
\end{document}